\documentclass[aip,cha,11pt,superscriptaddress,showpacs,floatfix,amsmath,amssymb,longbibliography]{revtex4-1}
\usepackage{color,graphicx,tabularx,bm,epstopdf,epsfig}

\begin{document}

\title{Accumulated densities of sedimenting particles in turbulent flows}

\author{Alessandro Sozza}
\email{asozza.ph@gmail.com}
\affiliation{Istituto dei Sistemi Complessi, CNR, via dei Taurini 19, 00185 Rome, Italy 
and INFN sez. Roma2 "Tor Vergata"}

\author{Gabor Drotos}
\affiliation{IFISC (CSIC-UIB), Institute for Cross-Disciplinary
Physics and Complex Systems, Campus Universitat de les Illes
Balears, 07122 Palma de Mallorca, Spain.}

\author{Emilio Hern\'andez-Garc\'ia}
\affiliation{IFISC (CSIC-UIB), Institute for Cross-Disciplinary
Physics and Complex Systems, Campus Universitat de les Illes
Balears, 07122 Palma de Mallorca, Spain.}

\author{Crist\'obal L\'opez}
\affiliation{IFISC (CSIC-UIB), Institute for Cross-Disciplinary
Physics and Complex Systems, Campus Universitat de les Illes
Balears, 07122 Palma de Mallorca, Spain.}

\begin{abstract}
We study the effect of turbulence on a sedimenting layer of particles 
by means of direct numerical simulations. A Lagrangian model in 
which particles are considered as tracers with an additional downward 
settling velocity is integrated together with an isotropic homogeneous 
turbulent flow. We study the spatial distribution of particles when they 
are collected on a plane at non-asymptotic times. We relate the resulting 
coarse-grained particle density to the history of the stretching rate along
the particle trajectory and the projection of the density onto the 
accumulation plane, and analyse the deviation from homogeneity 
in terms of the Reynolds number and the settling velocity.
We identify two regimes that arise during the early and during 
the well-mixed stage of advection. In the former regime, 
more inhomogeneity in the particle distribution is introduced for 
decreasing settling velocity or increasing Reynolds number, 
while the tendencies are opposite in the latter regime. 
A resonant-like crossover is found between these two regimes, 
where inhomogeneity is maximal.
\end{abstract}

\maketitle

\section{Introduction}
\label{sec:intro}

Sedimentation of particles in a turbulent flow is a crucial problem both for theory and applications. 
For example, it plays a key role in the process of rain formation in clouds \cite{falkovich2002,woittiez2009}. 
In the marine environment, sinking of particles is an important mechanism for many physical processes: 
in the sequestration of carbon dioxide \cite{delarocha2007,devries2012}, in the downward transport of 
organic and inorganic aggregates, such as marine snow \cite{alldredge1988,borgnino2019}, 
larval eggs and microplastics \cite{woodall2014,khatmullina2017}.
Experimentally, a way to estimate the downward fluxes of particles in the ocean interior is performed 
by placing sediment traps \cite{siegel1997,buesseler2007,siegel2008}. 
An open question concerns the identification of the mechanisms that lead to the observed size 
and spatial distributions of particles which are collected at a given depth by the traps.

The interaction between particles and flow is determinant to establish the spatial distribution 
of particles\cite{balkovsky2001}. Advection of a homogeneous distribution of passive particles 
in an incompressible flow generally results in a homogeneous concentration of particles.
Deviations from homogeneity may arise from some type of compressibility, either in the flow itself 
or in the motion of particles. 
In this case, particle dynamics is restricted to a lower-dimensional or even fractal subspace. 
Some exemplary cases of this phenomenon are found in the motion of particles 
under significant inertial effects \cite{bec2003,falkovich2004,dejoan2013,bec2014},  
in gyrotactic algae \cite{delillo2014}, in the action of buoyancy that forces particles to relax 
to a specific isopycnal depth \cite{sozza2018}, or even confines them to move on a horizontal 
sheet \cite{depietro2015} or on a free surface \cite{boffetta2004}. 
Another situation arises when considering initially inhomogeneous distributions. 
In this case, even with passive tracers in incompressible flow it is possible to observe 
inhomogeneities at non-asymptotic time scales. Cuts or projections to a lower-dimensional
manifold can give rise to additional inhomogeneity in this case. 
Under complex flow acting for sufficiently long times, the particle distribution will generally 
recover homogeneity, but for the finite times characteristic of realistic situations 
(for example sedimentation in the ocean) distributions are far from this asymptotic limit.

In this paper, we investigate the dynamics of a sedimenting layer of particles 
under three-dimensional turbulence and discuss the role of the flow to create inhomogeneities. 
Particles initially distributed homogeneously on an upper plane are let to fall down in a turbulent 
flow and are collected at a lower accumulation plane. 
We will discuss how the final coarse-grained density of particles is related to the properties of the flow. 
Two contributions were identified from previous works: 
stretching of the particle layer and projection on the collecting surface. 
Differently from the previous works, in which large-scale oceanic simulations
\cite{monroy2017, taylor2018} or chaotic dynamical systems \cite{drotos2019} were used, 
our attention is focused on the small-scale inhomogeneities due to an isotropic homogeneous
turbulent flow. In Section \ref{sec:form} we formulate the numerical setup 
and introduce the main theoretical tools. Sect. \ref{sec:results} describes our results 
and discusses them, and our conclusions are summarized in Sect. \ref{sec:conclusion}.

\section{Formulation of the problem}
\label{sec:form}

We begin by considering a homogeneous and isotropic turbulent flow 
described by an incompressible velocity field ${\bf u}({\bf x},t)$ 
(e.g. $\nabla\cdot{\bf u}=0$) ruled by Navier-Stokes equations
\begin{equation}
\partial_t {\bf u} + {\bf u} \cdot {\bf \nabla} {\bf u} =
- {\bf \nabla}p + \nu \triangle {\bf u} + {\bf f},
\label{eq:3}
\end{equation}
where $p$ is the pressure, $\nu$ is the kinematic viscosity and 
${\bf f}$ is the mechanical random forcing with imposed energy input $\varepsilon$.
In the absence of forcing and viscosity the system conserves
energy $E = \frac{1}{2}\langle {\bf u}^2 \rangle$. When
forcing and viscosity are at work, a turbulent steady state can
be reached, where energy is conserved only in a statistical sense 
and transferred from large scales to small scales with a constant flux \cite{frisch1995}.
The energy input $\varepsilon$, together with the kinematic viscosity $\nu$,
defines the Kolmogorov microscales for the length $\eta = (\nu^3/\varepsilon)^{1/4}$,
time $\tau_\eta=(\nu/\varepsilon)^{1/2}$, velocity $u_\eta=(\nu \varepsilon)^{1/4}$
and acceleration $a_\eta = (\varepsilon^3/\nu)^{1/4}$.
These scales will be used to define dimensionless parameters.

We now discuss the equations of motion for the particles. We
consider small spherical particles of size $a$ and density
$\rho_p$ transported by the incompressible velocity field ${\bf
u}({\bf x},t)$. A standard modeling set-up is a simplified form
of the Maxey-Riley equations \cite{maxey1983} for the velocity
of the particles ${\bf v}$:
\begin{equation}
\dfrac{d {\bf v}}{dt} = \beta\dfrac{d {\bf u}}{dt} - \dfrac{{\bf v}-{\bf u}}{\tau_p} + (1-\beta){\bf g} \ ,
\end{equation}
where $\beta=3\rho_f/(2\rho_p+\rho_f)$ is the density contrast
($\rho_f$ is the density of the fluid), $\tau_p =
a^2/(3\beta\nu)$ is the Stokes relaxation time and ${\bf g}$ is
the gravitational acceleration. If the flow is turbulent, we can
define two dimensionless parameters, the Stokes number
$St=\tau_p/\tau_\eta$ and the Froude number $Fr=a_\eta/g$. 
In the limits $St \rightarrow 0$ and $Fr \rightarrow 0$ but such
that $St/Fr$ remains constant, we can neglect the inertial
effects without omitting the gravity term (since the settling
velocity is\cite{sozza2016,drotos2019,mathai2016,Balachandar2010} $v_s \propto St/Fr$) 
leading to the reduced first-order differential equation
\begin{equation}
{\bf v(t) } \equiv \dfrac{d{\bf X}(t)}{dt} = {\bf u}({\bf X}(t),t) - v_s \hat{{\bf z}}.
\label{eq:1}
\end{equation}
In this expression we neglect terms of first order in $St$.
Thus particles are ``tracers'' transported by the
incompressible velocity field ${\bf u}({\bf x},t)$
that additionally sink with a constant settling velocity $v_s=(1-\beta)g \tau_p$
along the vertical direction $z$ (characterized by the
$\hat{\bf z}$ unit vector).
The model defined by Eq. (\ref{eq:1}) has been largely studied in the
literature and in previous works on this specific subject
\cite{siegel2008,fouxon2012,sozza2016,monroy2017,drotos2019}.
We remark that the model is derived within the assumption that
particles are small, with particle Reynolds number $Re_p = v_s
a /\nu \ll 1$, and not interacting, so that each particle
evolves independently from the others.
Furthermore, imposing $St<1$ restricts the validity of the reduced model to settling velocities 
$v_s < (1-\beta)g\tau_\eta$.

We introduce a dimensionless settling parameter $\Phi= v_s/U$, with $U$
being the root mean square velocity $U=(2E/3)^{1/2}$.
Notice that for $\Phi \gg 1$ (i.e. $v_s \gg U$)
the motion of the particles is ballistic and
turbulence is reduced to a small perturbation. On the contrary,
when $\Phi \approx 1$ or $\Phi \ll 1$ trajectories are strongly
controlled by turbulence and a random-like motion arises.
The constraint $v_s < (1-\beta)g\tau_\eta$, ensuring $St<1$,
reads as $\Phi < 15^{1/4} (1-\beta) Fr^{-1} Re_\lambda^{-1/2}$
in dimensionless quantities (see the definition of $Re_\lambda$ in Section~\ref{sec:sim}).

At the initial time $t=0$ particles are homogeneously
released at random positions on a horizontal plane $z=L$, after
which they move following Eq. (\ref{eq:1}). In order to
investigate the evolution and deformation of the layer of
particles we need to calculate, among other quantities, the
local stretching rates along each particle trajectory. 
We introduce the Jacobian matrix $\mathbb{J}(t)$ describing
separation in time $\delta {\bf X}(t)$ of particle trajectories
initialized at an infinitesimal distance $\delta {\bf X}(0)$,
i.e.
\begin{equation}
\delta X_\alpha(t) = \sum_{\beta=1,2,3}\mathbb{J}_{\alpha\beta}(t) \delta
X_\beta(0),
\label{eq:separation}
\end{equation}
with
\begin{equation}
\mathbb{J}_{\alpha\beta}(t) =
\dfrac{\partial X_\alpha(t)}{\partial X_\beta(0)},
\label{eq:jacobian}
\end{equation}
Using the chain rule, the evolution
of $\mathbb{J}_{\alpha\beta}$ is given by
\begin{equation}
\dfrac{d}{dt} \mathbb{J}_{\alpha\beta}(t) =
\sum_{\gamma=1,2,3} \partial_\gamma u_\alpha({\bf X}(t),t)
~ \mathbb{J}_{\gamma\beta}(t),
\label{eq:2}
\end{equation}
where $\partial_\gamma u_\alpha ({\bf X}(t),t)$ is the
fluid velocity gradient measured at the position of the
particle that started at ${\bf X}(0)$. The initial condition is
$\mathbb{J}_{\alpha\beta}(0)=\delta_{\alpha\beta}$. Since
initially the particle surface is horizontal, the first and
second columns of the matrix $\mathbb{J}_{\alpha\beta}(t)$ give
at each time two vectors, ${\bf t}_1(t)$ and ${\bf t}_2(t)$,
tangent to that falling surface.

We are interested in quantifying the final
distribution of particles deposited on a horizontal plane at a
fixed depth, say $z=0$. At that plane we can define a particle
surface density $\rho({\bf x}_{h})$, with ${\bf x}_h=(x,y)$
denoting the horizontal components. The relationship between
the homogeneous density $\rho_0$ at the upper release plane and
the density $\rho({\bf x}_{h})$ at the lower collecting plane
is given by a \emph{total density factor} $F({\bf x}_h)$
defined by $\rho({\bf x}_h)/\rho_0 \equiv F({\bf x}_h)$. As
demonstrated in previous work \cite{drotos2019,monroy2019},
this total factor is the product of two contributions: $F({\bf
x}_h)=S({\bf x}_h) P({\bf x}_h)$. $S$, the \emph{stretching
factor}, characterizes the stretching accumulated by the
falling surface around the trajectory that reaches the lower
plane at ${\bf x}_h$, whereas $P$, the \emph{projection
factor}, takes into account the orientation-dependent footprint
of the falling surface on the horizontal collecting plane in
the neighborhood of ${\bf x}_h$. These two factors can be
calculated \cite{drotos2019,monroy2019} (cf. \cite{pope1989,zheng2017} as well) 
from the tangent vectors ${\bf t}_1(t)$ and ${\bf t_2}(t)$ 
(and thus from Eq. (\ref{eq:2})) as
\begin{equation}
\left\{\begin{array}{l}
S = |{\bf t}_1 \times {\bf t}_2|^{-1}, \\[0.2cm]
P = \dfrac{\big| v_z \big|}{\big| {\bf \hat{n}}\cdot{\bf v} \big|}.
\end{array}
\right.
\label{eq:2a}
\end{equation}
All quantities are computed at the final time $t_h$ at which
the particle trajectory reaches position ${\bf x}_h$ on the
collecting plane. $v_z$ is the vertical component of the
particle velocity ${\bf v}$ at that time, and ${\bf \hat{n}}$
is the unit vector normal to the surface that can be computed
by normalizing ${\bf n}$, the vector normal to the surface
given by the cross product ${\bf n}(t_h)={\bf t}_1(t_h) \times
{\bf t}_2(t_h)$. If the falling surface reaches the
accumulation plane horizontally around ${\bf x}_h$, $\hat{{\bf
n}}$ at that location points along the $z$ axis and $P=1$,
meaning that there is no projection effect. Note that $P$
diverges where ${\bf \hat{n}}\cdot{\bf v} = 0$, i.e. where
particle velocity arrives at the collecting plane tangent to
the falling surface. These locations define \emph{caustics}
which form lines and typically occur when the falling surface
develops folds. On the other hand, the area of an infinitesimal
surface element at time $t$ is $|{\bf t}_1(t) \times {\bf
t}_2(t)|~dA_0$, where $dA_0$ is the initial area. Thus, $S=1$
if the surface reaches the accumulation plane unstretched.

\section{Numerical simulations}
\label{sec:sim}

We solve Eq.~(\ref{eq:3}) with a pseudo-spectral method on a
triply periodic cubic domain of size $L=2 \pi$ containing $M^3
= 32^3 - 256^3$ grid points to obtain statistically steady
flows with Taylor-microscale Reynolds number $Re_{\lambda} =
U\lambda/\nu \approx 19-93$, where $\lambda =
U\sqrt{15\nu/\varepsilon}$ is the Taylor microscale and $U$ is
the root-mean-square velocity fluctuation. Time marching is
performed using a second-order Runge-Kutta scheme. 
The forcing acts only at large scales in a shell of wavenumbers $k\leq k_f$,
and maintains a constant energy input $\langle {\bf f}\cdot{\bf u} \rangle = \varepsilon$,
which equates, on average, the energy dissipation rate.
This is obtained by taking 
${\bf f}({\bf x},t) = \varepsilon \Theta(k_f - k) {\bf u}({\bf x},t)/2E_{k\le k_f}$, 
where $\Theta$ is the Heaviside step function and $E_{k\le k_f}$ 
the kinetic energy restricted to the wavenumbers smaller than $k_f$ \cite{lamorgese2005,rosales2005,weiss2019}. 
We ensure that small-scale fluid motion is well resolved by imposing the
Kolmogorov length scale $\eta=(\nu^3/\varepsilon)^{1/4}$ of the
resulting flow to be of the same order as our grid spacing,
$k_{max}\eta > 1.8$, where $k_{max}=M/3$. Table~\ref{table1}
reports the most important Eulerian parameters used in the
simulations. Additional numerical details are as in \cite{sozza2020}.

\begin{table}[t]
\centering
\begin{tabular}{ccccccccccc}
\hline \hline
$M$ & $\nu$ & $E$ & $U$ & $u_\eta$ & $\mathcal{T}$ & $\tau_\eta$ & $\mathcal{L}$ & $\lambda$ & $\eta$ & $Re_\lambda$ \\
\hline
$32$ & ~$4 \times 10^{-2}$ & $0.47$ & $0.56$ & $0.25$ & $4.65$ & $0.63$ & $3.17$ & $1.36$ & $0.16$ & $19$ \\
$64$ & ~$2 \times 10^{-2}$ & $0.52$ & $0.59$ & $0.21$ & $5.25$ & $0.45$ & $3.80$ & $1.02$ & $0.09$ & $30$ \\
$128$ & $7 \times 10^{-3}$ & $0.61$ & $0.64$ & $0.16$ & $6.11$ & $0.26$ & $4.77$ & $0.65$ & $0.04$ & $60$ \\
$256$ & $3 \times 10^{-3}$ & $0.62$ & $0.65$ & $0.13$ & $6.24$ & $0.17$ & $4.93$ & $0.33$ & $0.02$ & $93$ \\
\hline \hline
\end{tabular}
\caption{Parameters of the four turbulent flows used here:
resolution $M$, kinematic viscosity $\nu$, 
kinetic energy $E=\frac{1}{2}\langle {\bf u}^2 \rangle$, 
root mean square velocity $U=(2E/3)^{1/2}$, 
Kolmogorov velocity $u_\eta = (\nu\varepsilon)^{1/4}$, 
eddy turnover time $\mathcal{T}=E/\varepsilon$, 
Kolmogorov time scale $\tau_\eta = (\nu/\varepsilon)^{1/2}$, 
integral length scale $\mathcal{L}=E^{3/2}/\varepsilon$, 
Taylor microscale $\lambda = U(15 \nu/\varepsilon)^{1/2}$, 
Kolmogorov length scale $\eta=(\nu^3/\varepsilon)^{1/4}$, 
Taylor-microscale Reynolds number $Re_\lambda = U\lambda/\nu$. 
All the simulations are performed with energy dissipation rate $\varepsilon=0.1$ 
and domain size $L = 2\pi$.} \label{table1}
\end{table}

\begin{figure}[t!]
\centering
\includegraphics[width=0.75\textwidth]{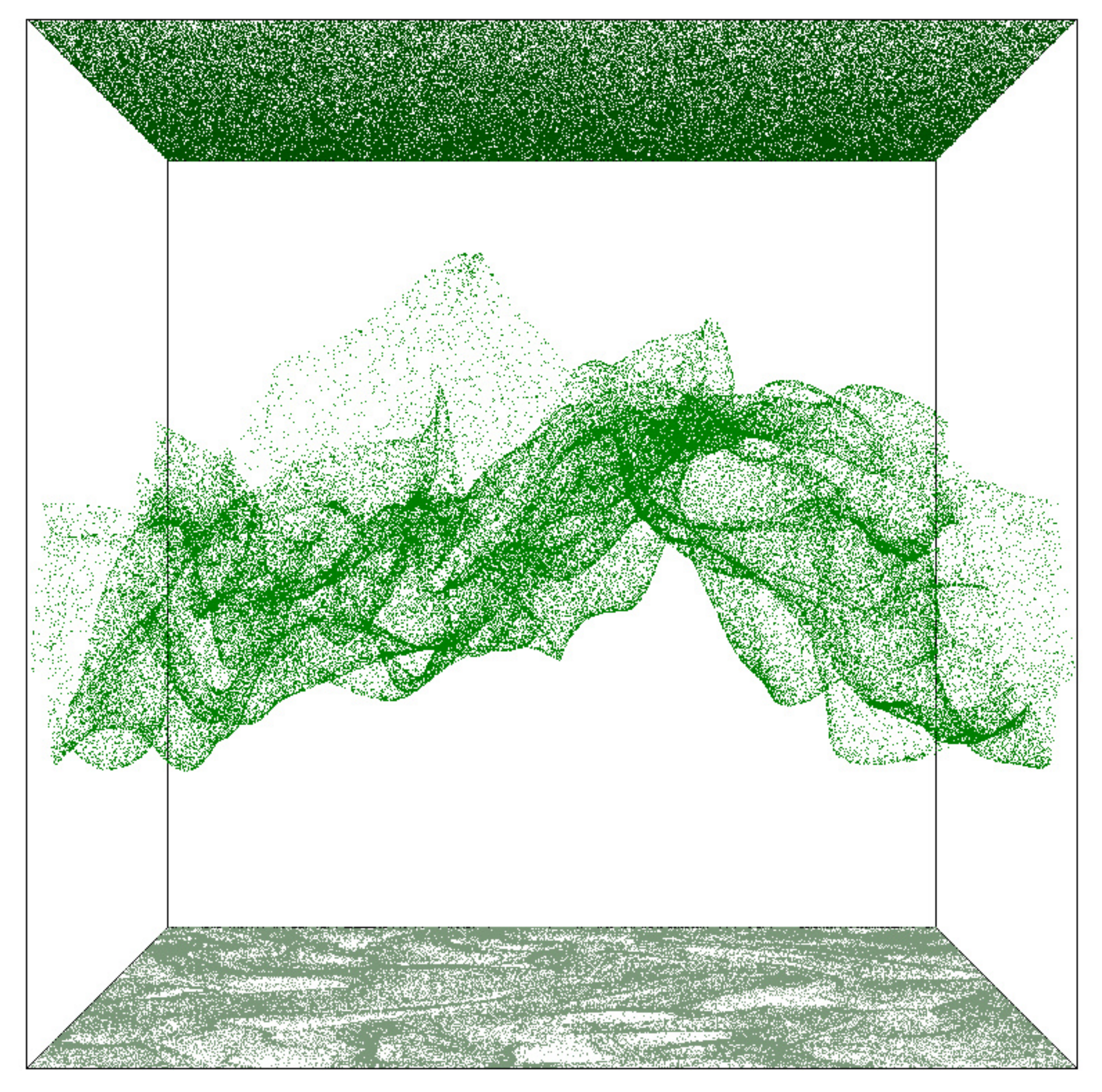}
\caption{Particle distribution in the initial homogeneous configuration at $z=L$ (upper plane), 
at an intermediate time (crumpled surface) and particles finally deposited at the lower plane $z=0$.}
\label{fig1}
\end{figure}

After the flow has reached statistical steady state,
$N = 1.2 \times 10^{6}$ particles are initialized with
homogeneously random positions on a plane at fixed horizontal
position $z_0=L$. The trajectory of each of them is evolved
with Eq. (\ref{eq:1}). The associated Jacobian matrix
$\mathbb{J}_{\alpha\beta}(t)$ giving deformations close to that
trajectory is simultaneously evolved with Eq. (\ref{eq:2}) and
initial condition
$\mathbb{J}_{\alpha\beta}(0)=\delta_{\alpha\beta}$. Fluid
velocity and its gradients are calculated by third-order
spatial interpolation on the particles' positions. The
integration time step $dt$ is chosen to be smaller than the
time needed to cross a grid cell, which is equal to satisfying 
the condition $v_s dt/dx < 1$, where $dx = L/M$. Deformation of
the evolving surface is characterized by its tangent vectors
${\bf t}_1(t)$ and ${\bf t}_2(t)$, given by the first two
columns of $\mathbb{J}_{\alpha\beta}(t)$, and by the normal
vector ${\bf n}(t)={\bf t}_1(t) \times {\bf t}_2(t)$. To limit
numerical errors arising from exponentially different values of
the components of $\mathbb{J}_{\alpha\beta}(t)$ a Gram-Schmidt
orthonormalization is applied periodically to the vectors ${\bf
t}_1(t)$, ${\bf t}_2(t)$ and ${\bf n}(t)$ and a new initial
condition for $\mathbb{J}_{\alpha\beta}$ is built by using the
resulting vectors as columns. The stretching factor $S$ in Eq.
(\ref{eq:2a}) is computed as a product of the partial
stretching factors obtained before each reinitialization.
We consider $17$ different values of the settling velocity
$v_s$. The largest values do not satisfy the constraint imposed by 
$St<1$ (a validity condition for the model, see section \ref{sec:form}), 
but $St = 1$ is clearly marked in every figure.

As we let particles fall and be transported by the flow, we
observe the deformation of the initially flat and homogeneous
distribution of particles into a crumpled surface, see
Fig.~\ref{fig1}. Since ${\bf \nabla}\cdot{\bf u}=0$ the
dynamics defined by Eq.~(\ref{eq:1}) is also incompressible
(${\bf \nabla}\cdot{\bf v}=0$), and we expect that a
homogeneous distribution (in the three-dimensional space) is
recovered after a sufficient number of eddy turnover times.
Such a return to homogeneity can be obtained either at large
times or, equivalently, at large depths. At finite times or
depths, we suggest that the settling parameter $\Phi=v_s/U$
determines the morphology of the surface.

Integration of particle trajectories is performed
until the particles reach the bottom plane at $z = 0$. In
principle, there may be particles that are trapped forever in
the flow above the bottom plane, but for the parameters used
here all particles arrive at the bottom plane within a finite
time. When a particle reaches the bottom plane at $z=0$, we
register its position ${\bf X}(t_h)=({\bf x}_h,0)$, its
velocity ${\bf v}$ and its arrival time $t_h$. With this
information and the values of the stretching computed along the
trajectory we are able to compute the total stretching $S$, the
projection $P$ and the total factor $F$ for each particle.

We recall that the simulations of the fluid dynamics are
implemented with periodic boundaries, which means that the
accumulation plane is neither a physical barrier nor a wall.
For the particles, however, the domain is periodic only in the
horizontal directions. In the vertical direction, it is
semi-finite with an absorbing boundary condition at the bottom,
on the accumulation plane, where particle trajectory
integration is stopped. We also remark that caution should be
taken when considering fast settling particles in a periodic
flow \cite{woittiez2009,ireland2016}, since they can perceive
spurious correlations of the turbulent flow, when the time it
takes a particle to fall through the domain is smaller than the
correlation time of the underlying flow. In standard setups
such as \cite{ireland2016}, particles were recirculating along
the periodic domain and then, if falling sufficiently fast,
they could artificially encounter the same eddy several times. In
our setup a fast particle can sample parts of the same eddy
twice at most. Also, the density factor accumulates stretching
contributions from the whole particle trajectory, of which the
boundary region is just a tiny part. Thus, we expect the
results described below to be independent of the use of
periodic boundary conditions. In fact we have computed the
average time-dependent stretching on particles with
trajectories stopped at the same accumulation layer, but with a
domain size $L$ for the flow simulation twice as large, and
found no difference with the result under the setup described
here.

\section{Results and discussion}
\label{sec:results}

\subsection{Direct inspection of spatial variations}
\label{sec:direct}

Particles reach the bottom with different times of arrival. Hence,
neighboring particles on the accumulation plane may have visited
different regions of the domain, experienced very different histories 
of stretching and folding and finally be collected at different moments.
Similarly, particles that are initially close may have diverged and concluded
their trajectories in very distant regions and at very diverse times as well.

First we aim to obtain a direct quantitative insight
to the inhomogeneities in the distribution of particles
collected on the accumulation plane. A suitable way to
characterize this concentration field is to compute a
coarse-grained surface density $\rho_{ij}$, where the indices 
$(i,j)$ label a set of boxes on the collecting plane: particle 
positions on that collecting plane are located within a
two-dimensional grid with resolution $M_B$ and counted in each
cell of size $L_B = L/M_B$. So that, $\rho_{ij} =
n_{ij}/L_B^2$, where $n_{ij}$ is the number of particles in the
cell $(i,j)$. Summing over all cells one obtains the total
number of particles as $\sum_{i,j=1}^{M_B} n_{ij} = N$.
The initial density, namely $\rho_0$, is equal to $N/L^2$, so
that $\rho_{ij}/\rho_0 = (n_{ij} L^2)/(N L_B^2) = (n_{ij}
M_B^2)/N$. In the homogeneous case when $n_{ij} = N (L_B/L)^2$,
we obtain $\rho_{ij}/\rho_0 = 1$. If the particle distribution
becomes inhomogeneous, the presence of voids and clusters will
be registered where $\rho_{ij}/\rho_0 < 1$ and
$\rho_{ij}/\rho_0 > 1$, respectively.

In the absence of folds, $\rho_{ij}/\rho_0$ is a
coarse-grained version of $F({\bf x}_h)=\rho({\bf
x}_h)/\rho_0$. If more than one branch of the surface appears
at a particular position due to some folding of the surface,
$\rho/\rho_0$ will correspond to a sum of the coarse-grained
values of $F$ characterizing the different branches.

\begin{figure}[t!]
\centering
\includegraphics[width=\textwidth]{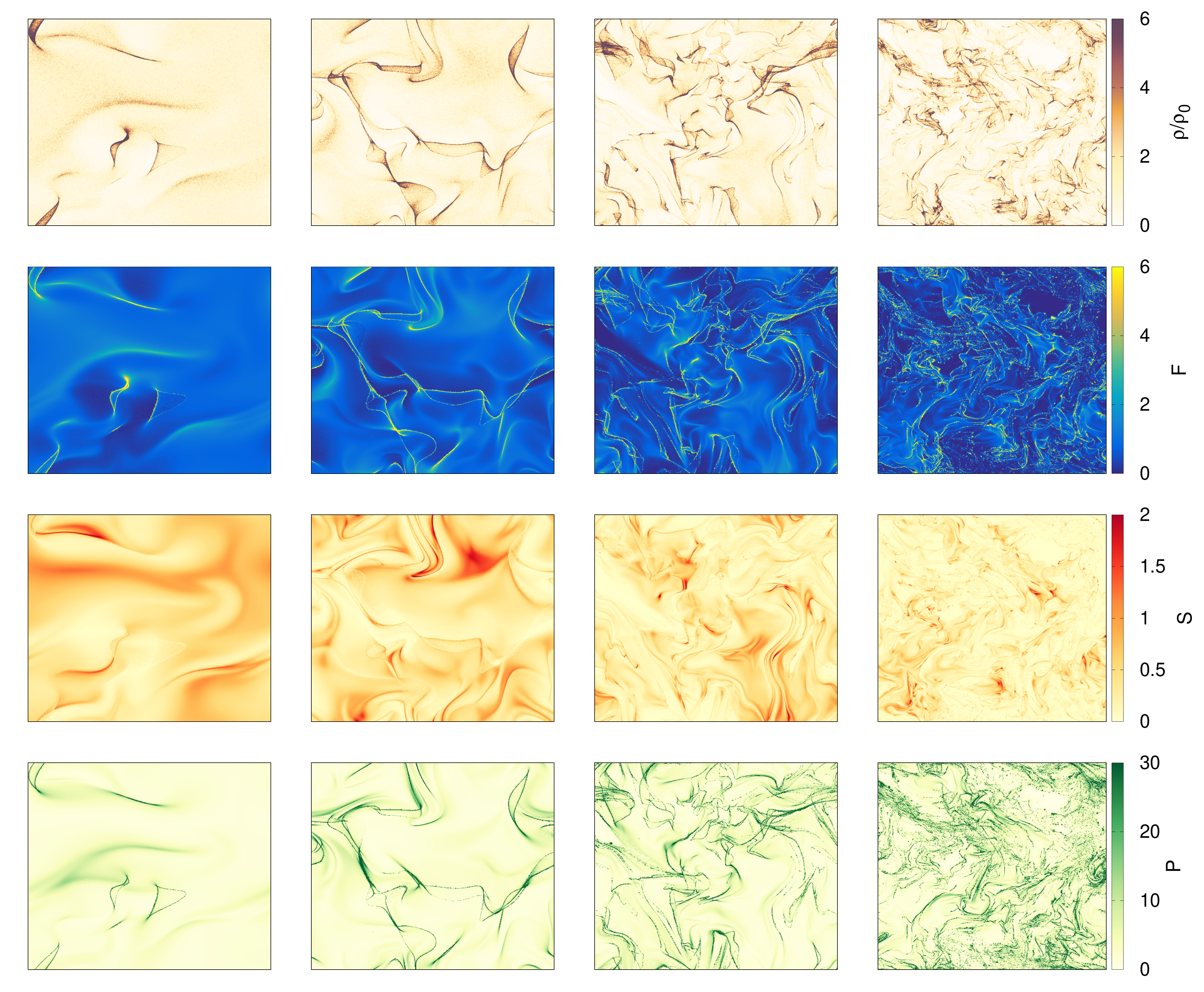}
\caption{Color map of the final distribution at the accumulation plane
of the coarse-grained particle density $\rho/\rho_0$
(first row, box resolution of the coarse-graining is $M_B=512$),
the total density factor $F({\bf x}_h)$ (second row) and the
separated contributions due to the stretching $S({\bf x}_h)$
and the projection $P({\bf x}_h)$ factors (third and fourth row).
$Re_\lambda= 19, 30, 60, 93$ in the four columns from left to right.
Computations are for $\Phi = 3$, a value for which a large Poisson
dispersion index $\chi$ is attained.}
\label{fig2}
\end{figure}

In Figure \ref{fig2}, examples for the spatial distribution of
the coarse-grained particle density, the total density factor,
and the separated contributions of stretching and projection
are shown on the accumulation plane for a given settling
parameter (chosen near the maximal observed inhomogeneity, as
characterized by the Poisson dispersion index $\chi$ defined
below). $F$, $S$ and $P$ have also been coarse-grained by
taking the arithmetic average in the same cells as those that
define $\rho_{ij}$. (Note that summation over different
branches is not actually performed for this qualitative
inspection.) We observe the emergence of clustering of
particles in the coarse-grained density and in the total
density factor, which are in reasonable agreement with each
other, even if a perfect agreement is not expected, since the
presence of folds is obvious. At most points we observe that
$S<1$ meaning that the infinitesimal area $|{\bf t}_1 \times
{\bf t}_2|~dA_0$ has grown larger than the original $dA_0$.
Also, the most noticeable features in $P$ are large values that
arise from the lines at which ${\bf \hat{n}}\cdot{\bf v} \to
0$, i.e. from the caustic lines at which $P$ diverges. In fact,
a comparison with the maps of stretching and projection
suggests that the largest inhomogeneities are due to the
formation of caustics, the abundance of which increases with
the Reynolds number $Re_\lambda$, leading to the formation of a
complex web of filaments. The dominance of caustics is similar
to the case of advection of inertial heavy particles, but in
that case they arise from the compressibility of the particle
flow \cite{wilkinson2005,gustavsson2014}, whereas here the
particle flow is incompressible ($\nabla\cdot {\bf v}=0$) and
develops caustics because of the two-dimensional character of
the initial distribution, together with the bending action of
the flow and the projection effect on the bottom surface. These
three effects concur in the formation of the final distribution of particles.

\subsection{Statistical characterization of inhomogeneities in the collecting plane}
\label{sec:statrho}

Next, we quantitatively investigate the degree of
inhomogeneity and its dependence on $\Phi$ and $Re_\lambda$ by
evaluating the so-called Poisson dispersion index $\chi$ of the
particle number distribution $n_{ij}$ defined over the
coarse-graining boxes of the accumulation plane. We also
discuss implications of the choice of the box size $L_B$ for
coarse-graining.

As a first step, the average and the standard
deviation of the set of values $\{n_{ij}\}$ on the
accumulation plane are considered (similarly as in
\cite{drotos2019,monroy2019}). Since the number of particles is
conserved and periodic boundary conditions are prescribed in
the horizontal direction, the spatial average of $n_{ij}$ is
the same as the initial number: $\overline{n_{ij}} = n_0 = N/M_B^2$ 
(where the bar represents the average with respect to boxes). 
Simple quantifiers of inhomogeneity are the standard
deviation $\sigma_n$ and its square, the variance. The
latter is conveniently normalized by $n_0$ to quantify
deviations from a homogeneous Poisson distribution by the
Poisson dispersion index as $\chi = \sigma_n^2/ \overline{n} = \sigma_n^2/n_0 $. 
Note that $\chi = 1$ corresponds to a homogeneous but random distribution,
describing particles arriving at uniformly random positions on
the accumulation plane. In such a case, a nonzero standard
deviation $\sigma_n$ results from the finite number of
particles, which, after coarse-graining, leads to a Poisson
distribution of $n$ over the boxes. True inhomogeneity, with
clusters and voids, is indicated by $\chi \neq 1$.

\begin{figure}[t!]
\centering
\includegraphics[width=\textwidth]{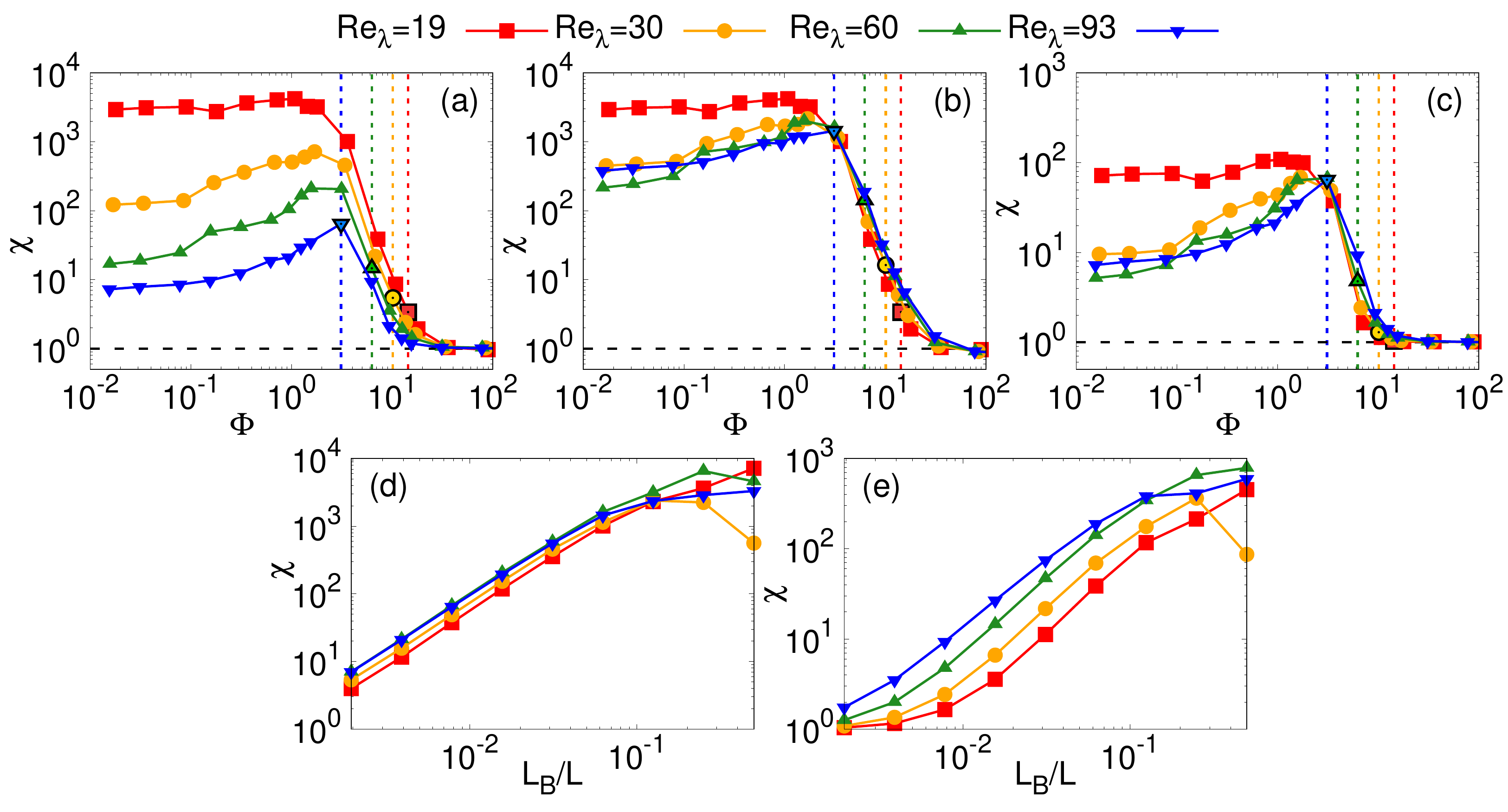}
\caption{The Poisson dispersion index $\chi$ computed
for the two-dimensional horizontal distribution of particles
at the accumulation level as a function of the settling parameter $\Phi$ 
(a-c) and the coarse-graining box size $L_B$ (d-e) for different values of $Re_\lambda$.
$L_B = 2L/M$ (a), $L_B = L/16$ (b) and $L_B = L/128$ (c), $\Phi = 3.35$ (d), $\Phi = 6.7$ (e). 
The settling parameter $\Phi$ corresponding to $St = 1$, 
an upper bound of the validity range of \ref{eq:1}, is marked by 
a black contoured symbol and a dashed line for each $Re_\lambda$.}
\label{fig3}
\end{figure}

How to choose $L_B$ for coarse-graining is not obvious. 
On the one hand, it is not meaningful to take $L_B$ below 
some mean distance between the particles ($\rho^{-1/2}$). 
On the other hand, $L_B$ may be chosen below the spatial resolution $L/M$ 
of the fluid flow in order to resolve small-scale folds of the particle sheet, 
which may have an important effect on the observed inhomogeneity. 
In Fig.~\ref{fig3}a) we present the dispersion index as a function
of the settling parameter $\Phi$, and where the size of the
coarse-graining boxes is chosen to depend on the resolution $M$
of the fluid model as $L_B = 2L/M$ and thus also on the
Reynolds number, cf. Table~\ref{table1}. This box size is near
the smallest characteristic scale (the Kolmogorov length scale)
of the fluid motion, but varies between relatively coarse
($L/16$) and much finer ($L/128$) values compared to the domain
size. Irrespective of $Re_\lambda$, particles are found
uniformly distributed on the accumulation plane for large
$\Phi$ ($\chi \approx 1$), which is a result of the lack of
time for the surface to deform (remember that the surface is
represented by randomly initialized particles). At intermediate
$\Phi$ we start to observe considerable inhomogeneities
characterized by $\chi > 1$.
A maximum of clustering is found between $\Phi = 1$ and $4$,
when the particle settling velocity $v_s$ is of the same order
as the root-mean-square fluid velocity $U$. Note also that the
accumulation plane would be reached during one unit of the
integral time scale $T$ by a particle uniformly settling with
$\Phi$ between $1.5$ and $2.5$ in all simulations, see
Table~\ref{table1}. Decreasing $\Phi$ further results in a
slight decrease of $\chi$.

Fig.~\ref{fig3}a also shows that
the limiting value of $\chi$ for $\Phi \to 0$ strongly depends
on the Reynolds number. For any $\Phi$, in fact, a higher
$Re_\lambda$ implies a smaller $\chi$. This result means that
inhomogeneities at the Kolmogorov length scale are actually
attenuated as the velocity field becomes increasingly
complicated, which can be attributed to an increased mixing.

One may, of course, also compare inhomogeneities
observed at the same spatial resolution $L_B$ in flows with
different Kolmogorov scale and $Re_\lambda$. Results are shown
for a large and a small $L_B$ in Figs.~\ref{fig3}b
and~\ref{fig3}c, respectively. While the characteristics of the
individual lines are the same as in Fig.~\ref{fig3}a), curves
for different $Re_\lambda$ cross at a value of $\Phi$ a bit
above $\Phi = 1$. That is, it depends on the settling velocity
whether increasing turbulence strength attenuates or enhances
inhomogeneity observed at a given spatial resolution. 
The settling parameter of Fig.~\ref{fig2} is just large enough 
to fall into the latter category.

It is worth noting that inhomogeneities observed at a
small resolution $L_B$ are typically weaker than those at a larger
resolution for any given Reynolds number: compare the range of
$\chi$ between Figs.~\ref{fig3}b and~\ref{fig3}c, and see
Figs.~\ref{fig3}d and~\ref{fig3}e for a direct representation
for given (large) values of $\Phi$. On the finest spatial scales, where
for fast settling initial randomness dominates over later
mixing, $\chi$ appears to converge to $1$.

We now see that the degree of observed inhomogeneity strongly 
depends on the spatial resolution, but its dependence on the settling velocity 
and on the turbulence strength ($Re_\lambda$) is robust for any given resolution. 
We can conclude about the existence of two regimes 
from the point of view of parameter dependence, 
one for large $\Phi$ and one for small $\Phi$, 
where the effect of increasing mixing by the flow is opposite. 
We will elaborate on this point and on the crossover 
between the two regimes in the next subsection, 
where we analyze the mechanisms underlying our observations.

When using the correlation dimension \cite{falkovich2002,bec2003,sozza2018} 
for estimating inhomogeneities as a function of $\Phi$ (not shown), 
the same qualitative behavior is observed as with the Poisson dispersion index.
This suggests that our conclusions are robust, and they do not depend on the choice 
of the particular statistical quantifier.

\subsection{Statistics of stretching and projection over trajectories}

We attempt to explore the
mechanisms leading to the dependence of $\chi$ on $\Phi$ and
$Re_\lambda$ presented in Fig.~\ref{fig3} by investigating
corresponding properties of the two mechanisms contributing to
inhomogeneities, namely the stretching and the projection
effects. For the statistical quantification of their local
characteristics, we treat different branches of the sedimenting
surface separately, without any summation. Furthermore, at
difference with Sect.~\ref{sec:statrho} and
\cite{drotos2019,monroy2019}, we explore in this section the
statistics with respect to the uniform distribution of
particles in the initial layer, or equivalently, 
we weight each particle trajectory equally. 
This provides a point of view complementary to the statistics 
over boxes in the collecting layer explored in Sect.~\ref{sec:statrho} 
to compute $\sigma_n$ and $\chi$. In particular, we compute here
arithmetic averages $\langle A \rangle$, standard deviations 
$\sigma_A$ and correlation coefficients of $A = S$, $P$ and
also $F$ over the individual values obtained for the individual particles, e.g.:
\begin{equation}
\langle A \rangle = \dfrac{1}{N} \sum_{k=1}^N A_k .
\end{equation}
where $k$ runs over different particles. In the limit of infinitely many particles,
\begin{align}
\langle A \rangle &=
\frac{ \int A({\bf x}_{0}) \,
\mathrm{d}^2{\bf x}_{0} }{ \int 1 \,\mathrm{d}^2{\bf x}_{0} } \nonumber \\
&= \frac{ \int A({\bf x}_h) F({\bf x}_h) \,\mathrm{d}^2{\bf x}_h }{ \int F({\bf x}_h) \,\mathrm{d}^2{\bf x}_h  } ,
\label{eq:Aaverage}
\end{align}
where the $d{\bf x}_{0}$ integrals are taken over the complete
initial release plane, and the integral over $d{\bf x}_h$ over
each branch of the surface sedimented on the collecting plane
with a subsequent summation of the values obtained for the
different branches. We have used that the number of particles
is conserved, $\rho_0 \mathrm{d}^2{\bf x}_{0} = \rho({\bf
x}_{h}) \mathrm{d}^2{\bf x}_{h}$. The second expression in
(\ref{eq:Aaverage}) illustrates why such a uniform weighting
according to the initial (uniform) distribution of the
particles is equivalent to weighting the points in the
collecting plane with the total density factor $F$ (or the
final density at those points if the sedimenting surface
reaches the collecting plane in a single branch). Note that
this kind of evaluation for a finite number $N$ of particles
corresponds to an "implicit" coarse-graining on the collecting
plane, with a grid provided by the particles' positions.

\begin{figure}[t!]
\centering
\includegraphics[width=\textwidth]{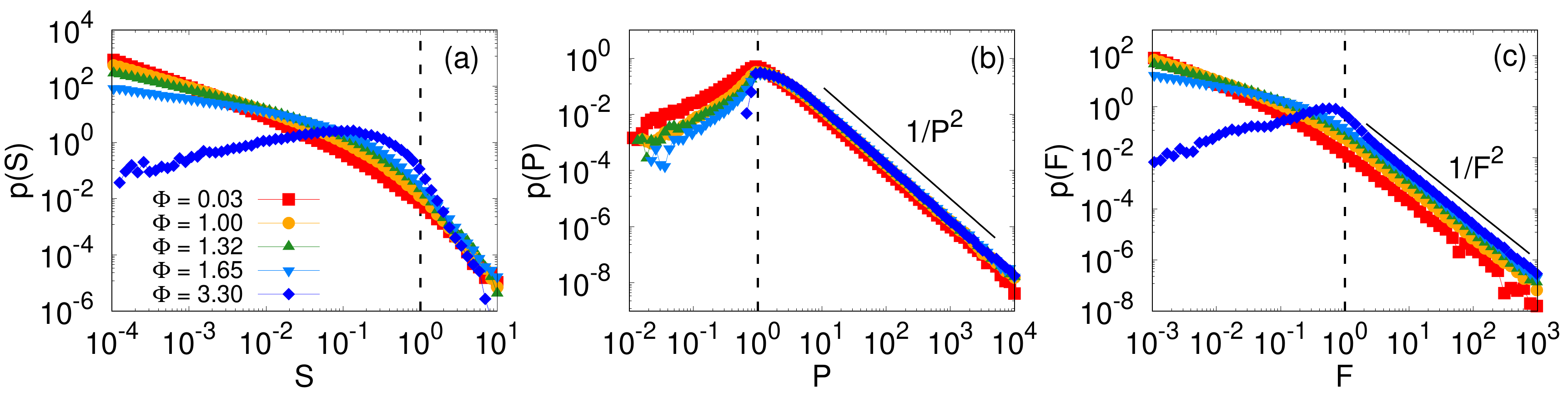}
\caption{Probability density functions (statistics over released particles) 
of stretching $p(S)$ (a), of projection $p(P)$ (b) and of the total density factor $p(F)$ (c), 
at $Re_\lambda=60$ for the indicated values of the settling parameter $\Phi$.}
\label{fig4}
\end{figure}

To better understand the contribution of stretching
and projection to the inhomogeneities, we first report in
Figure~\ref{fig4} the probability density functions of $S$,
$P$, and $F$ over the individual trajectories. 
The distribution of $F$ combines the behavior of $S$ and $P$.
The low values of the total density factor $F \ll 1$ are controlled by low
values of stretching, whereas large values $F \gg 1$ are
produced by the large values in $P$, associated with caustics.
The distributions of stretching appear to behave as power laws 
for small values of the settling parameter $\Phi$. 
When $\Phi$ is below roughly $1$ 
(the value giving the maximum of clustering, see Figs.~\ref{fig3}a-c), 
the weight given to very small values of $S$ increases as $\Phi$ decreases, 
since the areas of the surface elements arriving on the collecting plane 
can grow without limits.
On the contrary, the distribution of $P$ remains mostly unchanged for varying values of $\Phi$ 
and does not depend on $Re_\lambda$ (not shown), 
revealing a universal geometric feature of the projection near caustics. 
Indeed in Figure~\ref{fig4} we observe $p(P) \sim P^{-2}$ and
$p(F) \sim F^{-2}$ for $P \gg 1$ and $F \gg 1$, respectively,
which can be explained by the formation of caustics. It is a
well-known result that, generically, the density profile at a
line caustic diverges as $F \propto x^{-1/2}$, where $x$ is the
transverse spatial distance to the caustic \cite{wilkinson2005}.
Considering the transformation between variables $x$ and $F$
(assuming homogeneity in the direction parallel to the caustic),
$p(F) dF = p(x) dx$, and that, as seen before, the density factor
gives the proper weight to the horizontal locations $p(x)\sim F$
one obtains $p(F) = |dx/dF| p(x) \propto F^{-2}$.

\begin{figure}[t!]
\centering
\includegraphics[width=\textwidth]{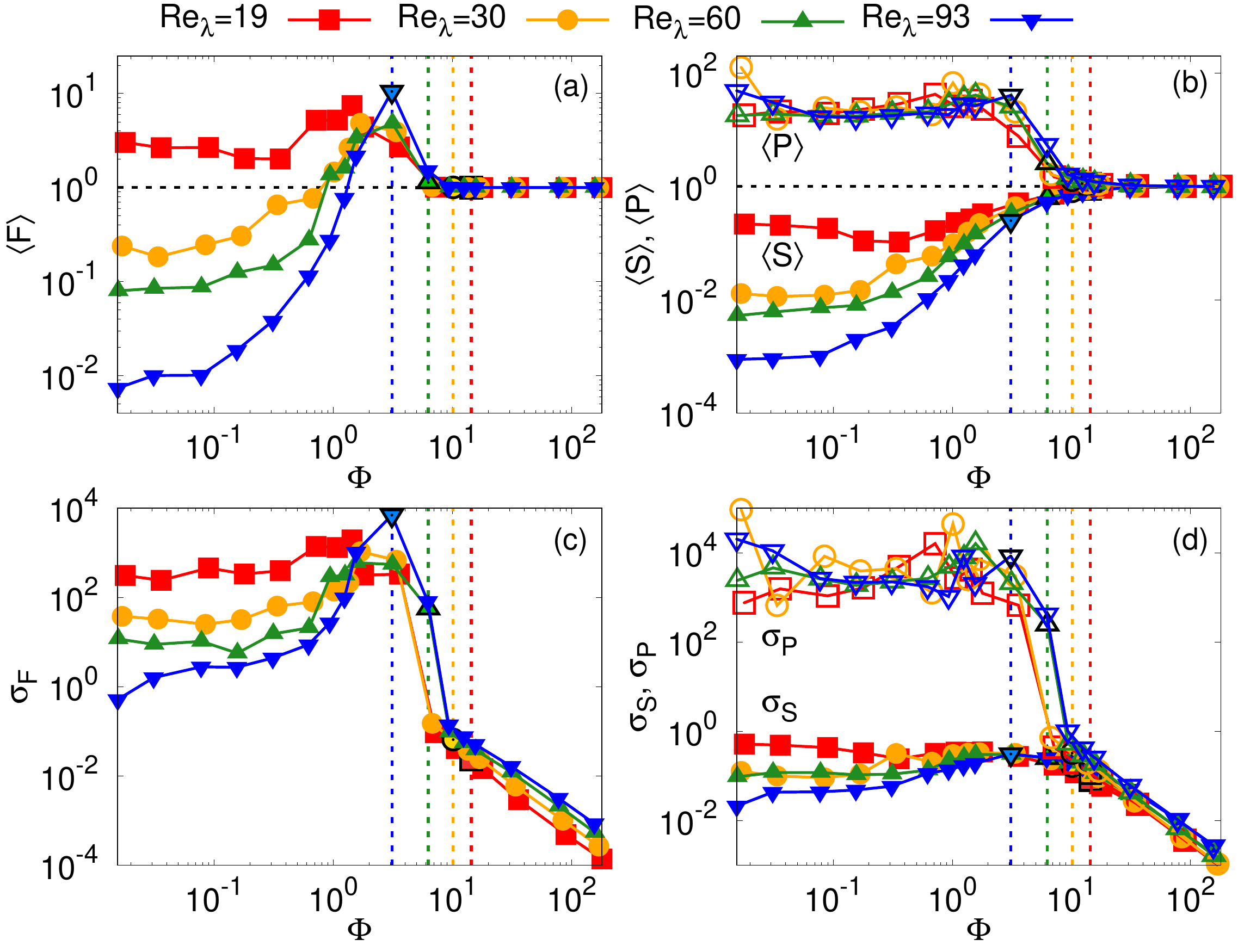}
\caption{Average and standard deviation of $F$, $S$ and $P$
for different values of $Re_\lambda$ as a function of the settling parameter $\Phi$. 
(a) Average of $F$, (b) average of $S$ and $P$, (c) standard deviation of $F$, 
(d) standard deviation of $S$ and $P$.
The settling parameter $\Phi$ corresponding to $St = 1$, 
an upper bound of the validity range of \ref{eq:1}, 
is marked by a black contoured symbol and a dashed line for each $Re_\lambda$.}
\label{fig5}
\end{figure}

To place the corresponding properties of $S$ and $P$ into a narrower context, 
we now investigate the average and standard deviation 
(statistics over released particles) of the total density factor $F$. 
$\langle F \rangle$ and $\sigma_F$ are plotted in Fig.~\ref{fig5}. 
The former characterizes the average dilution ($\langle F \rangle < 1$) 
or concentration ($\langle F \rangle > 1$) of particles 
on the collecting plane with respect to the initial release density $\rho_0$ 
(remember that different branches generated by folding of the falling surface 
are treated separately). Meanwhile, $\sigma_F$ describes the degree
of inhomogeneity among the different particles.

The shape of $\sigma_F$ as a function of $\Phi$ in Fig.~\ref{fig5} 
is very similar to that of $\chi$ in Fig.~\ref{fig3} 
(except for the large-$\Phi$ asymptotics, of course), 
which suggests that the accumulated inhomogeneities (represented by $\chi$) 
are closely related to the trajectory-wise processes of stretching and projection,
as opposed to summation of the density over different branches 
of the sedimented surface, which could also have a dominating effect. 
Note, however, that a quantitative comparison would be difficult, 
so that summation may well be important, too.

For increasing $\Phi$, $\langle F \rangle$
converges to $1$, as expected in the lack of time for
deformation and bending, while it generally exhibits a shift
toward net dilution, or area expansion, for decreasing $\Phi$
below $\Phi \approx 1$, which will be understood by analyzing
$S$ and $P$. Between $\Phi = 1$ and $10$, $\langle F \rangle$
exhibits a prominent maximum, just like $\chi$ and $\sigma_F$.
This maximum suggests again that either $v_s \approx U$ or a
settling time near the integral time scale (or both circumstances) 
result in a kind of resonance where maximal net deformation 
and maximal inhomogeneity in the deformation takes place.
This resonance represents, furthermore, a crossover between 
the regimes of large and small $\Phi$ with different tendencies.

Very close to $\Phi = 1$, just as for $\chi$, we find a crossing 
in the $Re_\lambda$-dependences, too: for $\Phi < 1$, 
an increasing Reynolds number results in a shift toward net 
dilution or expansion (decreasing $\langle F \rangle$) and a 
decrease in inhomogeneity ($\sigma_F$), but these tendencies
revert near $\Phi = 1$. We conclude that the effects of
increasing the strength of turbulence are far from trivial but
are certainly different in the regimes of small and large
settling parameter.

Much insight becomes accessible about explanations for the above
tendencies by analyzing the statistics of $S$ and $P$. The
decay of both $\sigma_S$ and $\sigma_P$ in Fig.~\ref{fig5}
follows the same power law for large $\Phi$ as $\sigma_F$.
$\sigma_S$ is not affected much by the resonance between 
$\Phi = 1$ and $10$, but typically becomes slightly decreasing for
$\Phi < 1$, while exhibiting only a minor degree of
inhomogeneity there. Based on this observation and the
similar magnitudes of $\sigma_F$ and $\sigma_P$ in
Figs.~\ref{fig5}, most of the inhomogeneity in $F$ near and
below the resonance might appear to originate from the
inhomogeneity of $P$. Note, however, the rather erratic behavior,
lacking a clear tendency, of $\sigma_P$ for decreasing $\Phi$,
contrasting the behavior of $\sigma_F$. While this relationship
will be further commented on later, and a comprehensive
understanding of all aspects is beyond the scope of the current
analysis, a universal conclusion about the small-$\Phi$
behavior of the degree of inhomogeneity in any quantity seems
to be a convergence to some constant value, in spite of the
arbitrarily long time available for deformation for $\Phi \to 0$.

This behavior of the standard deviation appears to 
apply to mean values as well, as Fig.~\ref{fig5} illustrates
for $\langle S \rangle$ and $\langle P \rangle$. With $\langle
P \rangle$ being mostly constant for $\Phi < 1$, the decrease
in $\langle F \rangle$ can naturally be linked to the decrease
of $\langle S \rangle$ for decreasing $\Phi$ observed in this
regime in Fig.~\ref{fig5}. The decrease in $\langle S \rangle$
below $1$ actually describes a stretching (expansion,
corresponding to a dilution of the density) of increasing
strength, which is presumably related to the longer time
available for the development of deformation. The same effect
may underlie the even sharper response of $\langle P \rangle$
for decreasing $\Phi$ between $10$ and $1$, before the increase
of $\langle P \rangle$ saturates. The difference in the
sharpness and the saturation of $\langle P \rangle$ is what
gives rise to the resonance-like behavior in $\langle F
\rangle$, even though $F = SP$ only pointwise, and $\langle F
\rangle \neq \langle S \rangle \langle P \rangle$ in general
due to spatial correlations.

Fig.~\ref{fig5} also provides with the opportunity to study the effects 
of varying the Reynolds number. Both $\langle P \rangle$ and $\sigma_P$
depend weakly and irregularly on $Re_\lambda$ for $\Phi < 1$.
This might be regarded as an indication of a saturation in all
effects of projection, which cannot be enhanced further by
modifying the circumstances (Reynolds number and settling
parameter). The explanation of such a saturation might be the
reaching of a "maximal randomness" in the orientation of the normal vector of an
arbitrarily chosen point of the sedimenting
surface\cite{drotos2019}.

The fact that $\langle S \rangle$ does not saturate
but decreases with increasing $Re_\lambda$ in the same range of
$\Phi$ (Fig.~\ref{fig5}, similarly to $\langle F \rangle$)
suggests that a similar saturation is not reached in the
stretching, the net effect of which may grow without any limit, 
as shown by the distribution of $S$ in Fig~\ref{fig4}a.
The dependence of $\langle S \rangle$ (and $\langle F \rangle$)
on $Re_\lambda$ might simply be understood as stronger
deformation resulting from stronger turbulence. Especially in
view of this, explaining why inhomogeneity is attenuated with
increasing $Re_\lambda$ as indicated by $\sigma_S$ might be
linked to the long-term homogenization in an increasingly
complicated flow with increasing mixing capability. The
attenuation of inhomogeneity with decreasing $\Phi$ might be
explained in a similar way but relying on the longer time
available for mixing instead of the increasing mixing
capability of the flow.
How $\sigma_S$ depends on $Re_\lambda$ and $\Phi$ for $\Phi < 1$ 
appears to be transferred to $\sigma_F$ (Fig.~\ref{fig5}),
which suggests that inhomogeneities in stretching do have an 
important effect on the final inhomogeneities in spite of their 
much smaller magnitude. 

By now, mixing is understood to be a central process in shaping 
the inhomogeneities for $\Phi < 1$. We have seen that more mixing 
(smaller $\Phi$ or larger $Re_\lambda$) attenuates inhomogeneities 
on the long term (at least when investigated at a predefined spatial resolution, 
which is determined here by the finite number of particles, cf. Section~\ref{sec:statrho}). 
Without mixing, however, there would be no inhomogeneities at all.

We resolve this apparent contradiction by considering the short-term effects of mixing. 
In particular, when compared to the small-$\Phi$ regime,
mixing works in the opposite way in the large-$\Phi$ regime. 
That is, $\sigma_P$ and $\sigma_S$ 
(and also $\langle P \rangle$ and $\langle S \rangle$) 
increase with increasing $Re_\lambda$ (see in Fig.~\ref{fig5}). 
The presumable explanation precisely lies in the time available for mixing,
which is around or less than the integral time scale. It seems
plausible that saturation is not reached in the effects of 
the projection, nor homogenization is performed, which is 
confirmed by $\sigma_P$ and $\sigma_S$ growing from $0$ with 
decreasing $\Phi$ and increasing $Re_\lambda$ in Fig.~\ref{fig5}. 
As long as the sheet is not deformed very much, 
stronger turbulence or longer time naturally results in 
the intensification of both the net effects of deformation and 
their inhomogeneity. For the net effects $\langle P \rangle$ 
and $\langle S \rangle$, this is similar to the $\Phi < 1$ 
regime except that $\langle P \rangle$ saturates there.

Comparing the $Re_\lambda$-dependence of $\langle S \rangle$ and $\langle P \rangle$ 
for $\Phi \gtrsim 1$, the former becomes weaker than the latter, 
and this is what we suppose to yield a change in the dependence of $\langle F \rangle$ 
on $Re_\lambda$ between the two regimes. 
At the same time, the similar change for $\sigma_F$ is more straightforwardly explained 
by the same change for $\sigma_S$ and $\sigma_P$, 
corresponding to an inherent difference between the short-term and long-term behaviors.
It is interesting to observe that introducing stronger turbulence 
enhances and attenuates inhomogeneities before and after the crossover.

So far, we have learnt that observable inhomogeneities (as in Fig.~\ref{fig3}) 
are strongly determined by trajectory-wise processes (investigated in Fig.~\ref{fig4} and \ref{fig5}). 
Increasing mixing by the flow has been identified to introduce and enhance inhomogeneities 
on the short term, and to attenuate them on the long term 
(at the spatial resolution corresponding to the finite number of particles). 
We point out, however, that summation over the increasingly 
numerous branches of the falling layer may contribute to the attenuation of inhomogeneities. 
Irrespective of that, the strongest inhomogeneities have been linked to the projection of the falling layer 
onto the accumulation plane (e.g. caustics). 
However, the impacts of projection have been found to saturate after entering the well-mixed regime, 
where the parameter dependence of stretching effects appears to be dominant, 
conforming with the above-mentioned attenuation of observable inhomogeneities.

\begin{figure}[t!]
\centering
\includegraphics[width=\textwidth]{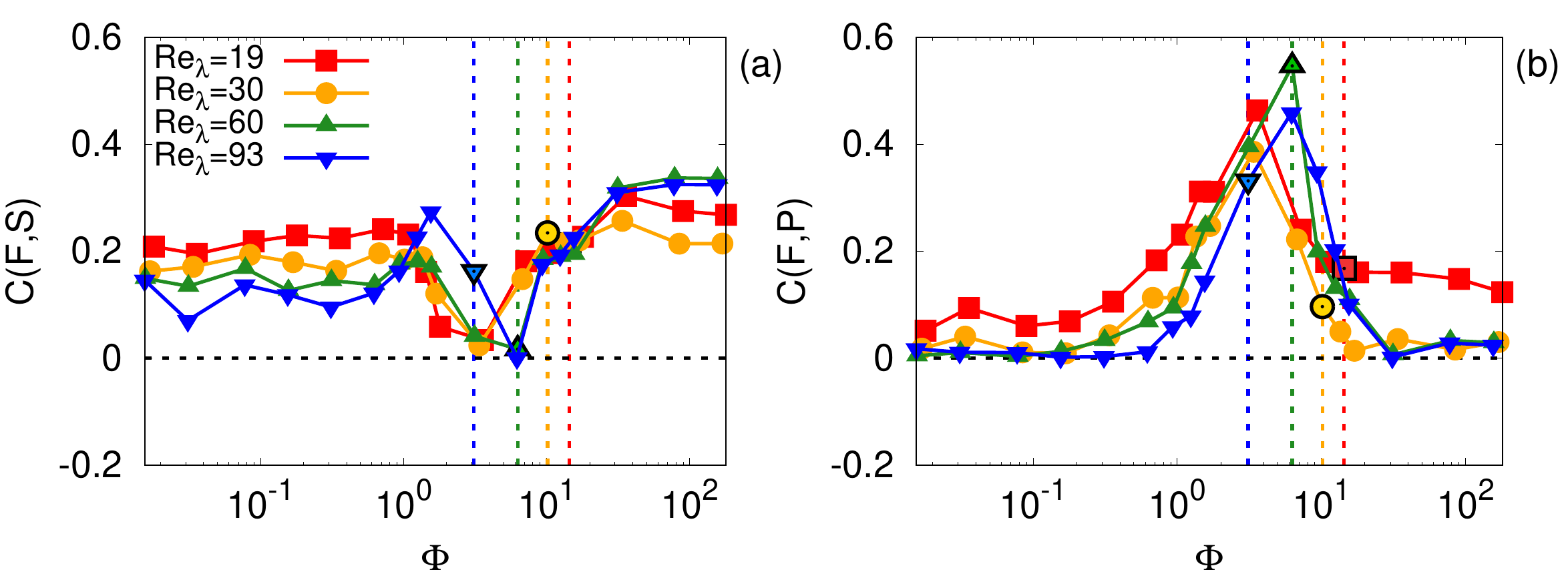}
\caption{Pearson correlation coefficients: (a) $C(F,S)$ and (b) $C(F,P)$ for
different values of $Re_\lambda$ as a function of the settling parameter $\Phi$. 
The settling parameter $\Phi$ corresponding to $St = 1$, 
an upper bound of the validity range of \ref{eq:1}, 
is marked by a black contoured symbol and a dashed line for each $Re_\lambda$.}
\label{fig6}
\end{figure}

The latter claims, indirectly derived from Fig.~\ref{fig5}, 
are supported by an analysis of spatial correlations.
We compute the Pearson correlation coefficient
of $F$ with $S$ and $P$ ($C(F,S)$ and $C(F,P)$) using
statistics over particles.
The value of the correlation coefficient is influenced by both
net effects and inhomogeneities. If stretching and projection
were uncorrelated (which is not the case, see Fig.~\ref{fig2}), 
we would have $C(F,S) = \frac{\sigma_S}{\sigma_F} \langle P \rangle$
and a corresponding formula for $C(F,P)$,
which suggests that both averages and standard deviations
are relevant \cite{monroy2019}.

Results for the correlation coefficients are plotted in Fig.~\ref{fig6}. 
For increasing $\Phi$ beyond the crossover, stretching seems to be 
dominant in forming the spatial structures of the final density
(although this is not observed for all Reynolds numbers).
This suggests that undulations of the surface become negligible
compared to the effect of stretching for increasing $\Phi$, 
in accordance with the same conclusion of \cite{monroy2019}.
In the vicinity of the resonance, projection takes over, and
the correlation with stretching falls to zero. 
This is presumably due to the increase in the inhomogeneity
of projection, without a similar increase for stretching (see Fig.~\ref{fig5}d).
Such a result is in agreement with the qualitative observation 
of the spatial distributions in the proximity of the resonance, 
displayed in Fig.~\ref{fig2}, where the filamentary structures 
of $P$ and $F$ appear to be well correlated. 
The dominance reverts again for $\Phi < 1$,
which is in accordance with the observation that both 
$\langle F \rangle$ and $\sigma_F$ follow the corresponding 
features of $S$ (both as a function of $\Phi$ and $Re_\lambda$). 
In relation with Fig.~\ref{fig5}, we explained this via the saturation of $P$, 
corresponding to the unit vector $\hat{{\bf n}}$ normal to the (wrapped and
contorted) surface taking already a random orientation, 
which cannot become more disordered by decreasing $\Phi$.

\begin{figure}[t!]
\centering
\includegraphics[width=\textwidth]{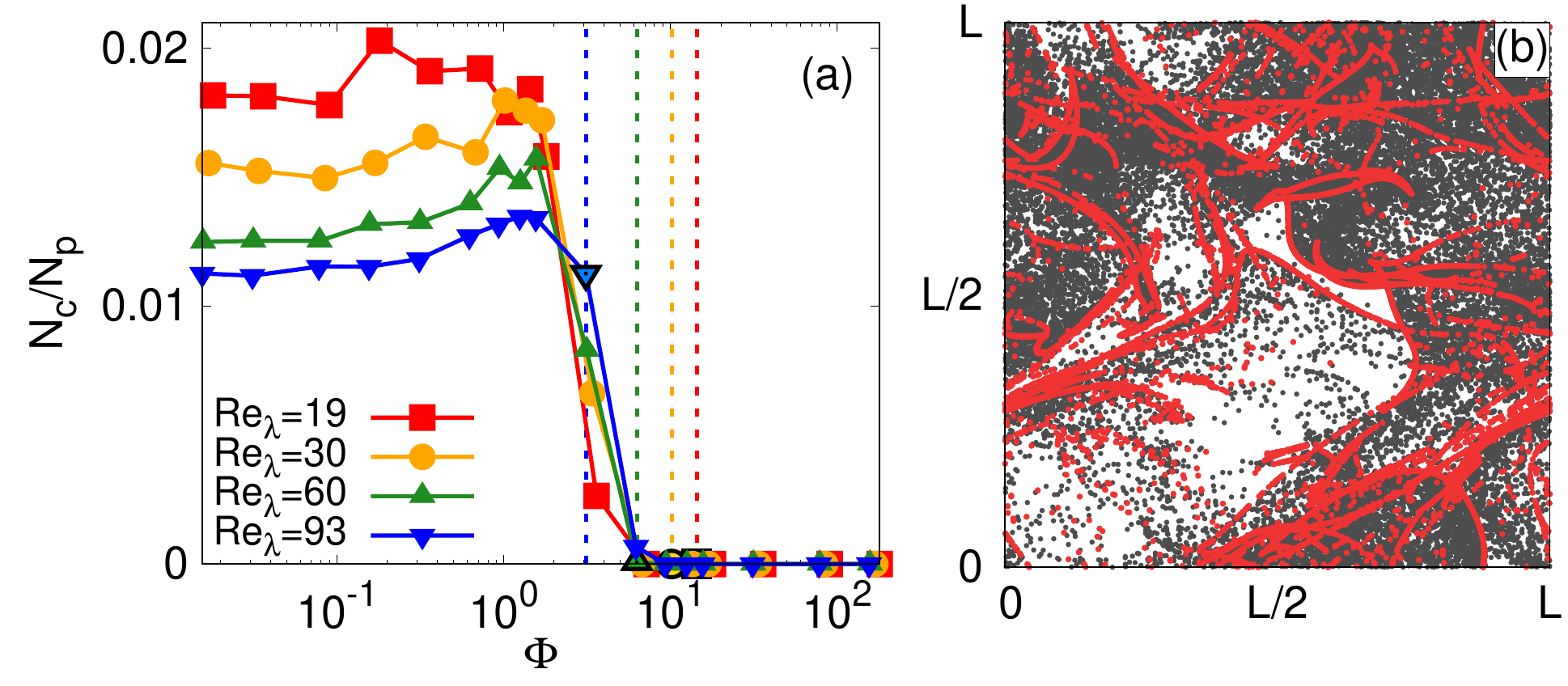}
\caption{(a) Fraction of particles $N_c/N_p$ in caustics characterized 
by having ${\bf \hat{n}}\cdot{\bf v}=0$ 
(numerically, the requirement is $|{\bf \hat{n}}\cdot{\bf v}|<0.01$). 
The settling parameter $\Phi$ corresponding to $St = 1$, 
an upper bound of the validity range of \ref{eq:1}, 
is marked by a black contoured symbol and a dashed line for each $Re_\lambda$. 
(b) Positions of the particles sedimented on the collecting plane. 
In red, particles in caustics as defined above. $\Phi=1$ and $Re_{\lambda}=19$. }
\label{fig7}
\end{figure}

In Fig.~\ref{fig7}, we present further evidence supporting that 
the reason for the huge increment in $P$ in vicinity of the resonance 
(coming from large $\Phi$ where the surface is flat, see Fig.~\ref{fig5}), 
is that caustics appear where the density formally diverges. 
Thus in Fig.~\ref{fig7}a we plot the fraction of particles in caustics 
(numerically requiring $|{\bf \hat{n}}\cdot{\bf v}|<0.01$) as a function of $\Phi$, 
observing the two regimes: very small for large values of $\Phi$, 
and non-negligible values for $\Phi < 1$ and any values of $Re_\lambda$. 
The example of Fig.~\ref{fig7}b illustrates by a direct plotting of the positions 
of the particles on the accumulation plane that caustics are closely related 
to inhomogeneities in the sedimented particles' distribution. 
The filamentary pattern of caustic lines in Fig.~\ref{fig7}b is recognized to be the same 
as that of the maxima of $P$ in Fig.~\ref{fig2} and spreads through the collecting plane.

\section{Conclusions}
\label{sec:conclusion}

We performed direct numerical simulations of sinking
non-inertial particles in a turbulent flow, exploring a range
of settling velocities and Reynolds numbers. We focused our
attention on the inhomogeneities of the particle distribution
that take place when particles released on a plane at a fixed
height are collected on a certain accumulation depth.

Although the Lagrangian dynamics is incompressible, 
advection of the two-dimensional surface by the flow 
and accumulation on a plane can lead to the emergence
of inhomogeneities by a combination of stretching and
projection effects \cite{drotos2019,monroy2019}. 
Our results indicate the existence of two different regimes from this point of view:
the inhomogeneities grow during the initial stages of the
dispersion, while they undergo attenuation when approaching the
long-term asymptotics of a well-mixed state.
With a fixed domain size, the settling time and thus the
degree of inhomogeneity in the accumulated density is
controlled by the settling velocity: the initial and the
long-term regimes are realized for large and small settling
velocity, respectively.

Between the two regimes, we have found a
"resonant" range of settling velocity where inhomogeneity can
become maximal. The maximum might approximately be determined
by the coincidence of the settling velocity with the
root-mean-square velocity of the flow, by the coincidence of
the typical settling time with the integral time scale of the
flow, or by an interplay of the two circumstances.

The range of settling velocity hosting this
resonance-like behavior not only marks a change of behavior of
the degree of inhomogeneity as a function of the settling
velocity itself, but also as a function of the Reynolds number.
During the initial transients, a more complicated flow (higher Reynolds number) enhances
inhomogeneity, while it facilitates approaching homogeneous
mixing in the regime leading to the long-term asymptotics.

We have also investigated the contributions 
of the two basic inhomogeneizing mechanisms in the two regimes. 
For large settling velocities, when the surface is
bended very little without developing overhangs, stretching is
predominant. When getting close to resonant-like settling
velocities, folds appear, yielding projection caustics in the
sedimented density. For this reason, effects of projection
become dominant, and this is responsible for the crossover in
some properties at the resonance-like region. With further
decrease of the settling velocity, the magnitude of the
inhomogeneities remains determined by projection, but the
increasing effects of projection saturate soon as mixing becomes strong. 
The parameter dependence of observable inhomogeneities then conforms 
with the increasing homogeneity of stretching as mixing becomes stronger, 
although summation over a large number of different branches 
of the falling particle layer may also be important.

The above results give an opportunity to comment on some previous work. 
Although our setup shows a few important differences
with the problem of sedimentation in mesoscale oceanic flows
addressed in \cite{monroy2019} common points are prominent enough 
to locate the mesoscale oceanic setup on the axis of the settling velocity.
In particular, although anisotropy in the velocity field of the
ocean is pronounced (with large differences between
horizontal and vertical velocities), one can safely state that
the settling velocity of typical biogenic particles\cite{monroy2017} 
is (several times) larger than vertical velocities of flow. As for the 
typical sedimentation time, it is the same order of
magnitude as the characteristic time scale of the mesoscale
oceanic flow. These circumstances may mean that the parameters
are not far from the resonance-like maximum of inhomogeneity, and 
fall into the regime of initial transients identified for $\Phi > 1$ in the present paper. 
Considerable inhomogeneities appear in corresponding oceanic
simulations and are enhanced for decreasing settling velocity and 
increasing mesoscale turbulence strength \cite{monroy2019}.

Finally, we indicate the relevance of these studies for sinking biogenic particles 
in an eddy-resolving oceanic velocity field. A careful study has been performed in
\cite{monroy2019} for a mesoscale oceanic flow, 
based on the analysis of \cite{monroy2017} of sizes and
densities of particles for which our modeling approach is valid. 
These biogenic particles (examples of which are dead plankton bodies, 
zoo-plankton fecal pellets, or small aggregates and marine snow) 
have typical sizes $a$ ranging between $10^{-6} - 10^{-3} ~ \mbox{m}$, 
and typical densities between between $\rho_p = 1050-2700 ~ \mbox{kg} \mbox{m}^{-3}$, 
so that $\beta$ is bounded between $0.5$ and $1.0$. 
Oceanic turbulence is characterized by 
$\varepsilon = 10^{-4} - 10^{-8} ~ \mbox{m}^2/\mbox{s}^3$,
$\nu=10^{-6} ~ \mbox{m}^2/\mbox{s}$ \cite{Thorpe2007}, 
for which we obtain a Kolmogorov length scale 
$\eta = (\nu^3/\varepsilon)^1/4 = 0.3-3 ~ \mbox{mm}$, 
Kolmogorov time scale $\tau_\eta=(\nu/\varepsilon)^{1/2}=0.1-10 ~ \mbox{s}$,
Kolmogorov velocity $u_\eta = (\nu \varepsilon)^{1/4} = 0.3-3 ~ \mbox{mm}/\mbox{s}$ 
and acceleration $a_\eta = (\varepsilon^3/\nu)^{1/4} = 30 - 0.03 ~ \mbox{mm}/\mbox{s}^2 $,
leading to a Froude number $Fr=\times 10^{-6} - \times 10^{-3}$ 
and a Stokes number $St=10^{-7} - 0.5$. 
The range of Reynolds numbers used in 
our numerical simulations indicate that we are dealing with 
spatial scales of the flow between $6 \ cm$ and $1 \ m$. 
Note that our configuration represents two relevant situations in this
context: one is sedimentation on the seafloor, and the
other is the collection of particles in sediment traps located
at a given depth. For this last situation, the impact of
boundary conditions at the bottom is irrelevant. However, for
sedimentation on the seafloor, in the case of a no-slip boundary
condition, a boundary layer close to the bottom is formed and
turbulence is drastically reduced and modified there, 
which, does not affect the processes in the bulk \cite{monroy2019}.

\section*{Acknowledgments}
AS acknowledges support from grant MODSS
(Monitoring of space debris based on intercontinental stereoscopic detection) 
ID 85-2017-14966, research project funded by \emph{Lazio Innova/Regione Lazio} 
according to Italian law L.R. 13/08. 
GD, EH-G and CL acknowledges support from the Maria de Maeztu Program 
for units of Excellence in R\&D ( MDM-2017-0711). 
GD also acknowledges support from the European Social Fund under CAIB grant 
PD/020/2018 "Margalida Comas", and from the Hungarian grant NKFI-124256 (NKFIH).
The data that support the findings of this study are available 
from the corresponding author upon reasonable request.

\bibliographystyle{ieeetr}
\bibliography{biblio_def.bib}

\end{document}